October 31, 2013

# 5 (Upgradable to 25 keV) Free Electron Laser (FEL) Facility

R. C. York

Michigan State University
East Lansing, MI 48824

## Abstract

A Free Electron Laser (FEL) facility utilizing a recirculated Superconducting Radio Frequency (SRF) linear accelerator (linac) provides the opportunity to achieve about five times greater photon energy than an unrecirculated linac of similar cost.

- A 4 GeV SRF, cw, electron linac can be used to drive an FEL producing 5 keV photons. The SLAC National Accelerator Laboratory, a Department of Energy (DOE) Basic Energy Sciences (BES) laboratory, proposes to utilize a 4 GeV unrecirculated, SRF, linac in a segment of existing linac tunnel.
- For an initial investment similar to that of the proposed SLAC strategy, a recirculated SRF linac system could deliver the 4 GeV electrons for photon energies of 5 keV *and provide an upgrade path to photon energies of 25 keV*.
- Further support amounting to about a third of the initial investment would provide upgrade funds for additional SRF linac and cryogenic capacity sufficient to provide electron energies appropriate for 25 keV photons matching the European XFEL.

## Introduction

In July 2013, the DOE BES Advisory Committee (BESAC) issued a report [1] recommending an FEL with photon energy of about 5 keV. A design concept for the electron beam accelerator suitable for a 5 keV facility is presented. With appropriate siting, an initial implementation strategy could be the realization of the 5 keV facility with an upgrade path to a world-class, scientifically important 25 keV cw facility. (See, for example, Reference [2].)

### FEL SRF Linac Design Issues

With a conventional planar magnetostatic undulator, the FEL process generates a photon wavelength $\lambda ph$ given by:

$$\lambda ph = [\lambda u \cdot (1 + K^2/2)]/(h \cdot 2 \cdot \gamma^2) \quad \textbf{Eqn. 1}$$

where $\gamma$ = (electron kinetic energy/electron mass + 1), $\lambda u$ = undulator period. For a magnetostatic undulator $K = 0.934 \cdot B(T) \cdot \lambda u(cm)$ with B = undulator on-axis magnetic field, and $h$ = harmonic (1,3,5…) of the photon radiation.

The envisioned FEL-based science program is best realized with a cw linac providing more stable and precise photon delivery at non-destructive levels compared to low duty factor linacs. The primary cost element for a cw FEL is the SRF linac. From Eqn. 1, $\gamma \propto \sqrt{\lambda u/\lambda ph}$. The necessary linac energy to achieve a specific photon energy can be reduced as $\sqrt{\lambda u}$, but for magneto-static undulators, $\lambda u$ have only been reliably developed to ~cm level with reasonable K values, which determine the FEL gain.

In addition, there is a beam quality requirement that the electron beam necessary to achieve optimal FEL performance must be correlated with $\lambda ph$ as:

$$\varepsilon_g = \varepsilon_N/\gamma \leq \lambda ph/(4 \cdot \pi) \quad \text{Eqn. 2}$$

where $\varepsilon_g$ = geometric emittance, $\varepsilon_N$ = normalized emittance.

Therefore, utilization of lower energy electrons has the associated requirement of a smaller $\varepsilon_N$ given the reduced $\gamma$.

The longitudinal beam parameter ($\Delta\gamma/\gamma$) has a similar beam quality requirement of

$$\Delta\gamma/\gamma < \rho \quad \text{Eqn. 3}$$

with:

$$\rho = (1/4\gamma)[(Ipk \cdot \lambda u^2 \cdot K^2 \cdot [JJ]^2)/(IA \cdot \pi^2 \cdot \varepsilon_g \cdot \beta_f)]^{1/3} \quad \text{Eqn. 4}$$

where $\rho$ = FEL Pierce parameter, $Ipk$ = peak beam current, $[JJ]$ = Bessel function factor for planar undulator, $IA$ = Alfen current, $\beta_f$ = machine beta function.

## Design Concept

The cost of providing a cw electron beam with the energy necessary for a given photon energy is driven by the SRF linac costs. A recirculated SRF linac reduces the amount of linac necessary for a given energy and therefore, the cost.

The design is based on the existing Jefferson Lab (JLab) 12 GeV upgrade, 1.5 GHz, cryomodule providing an energy gain of 100 MeV in a length of ~10 m including an intra-cryomodule warm region for diagnostics and transverse focusing. Though some improvements like e.g., intra-cell stiffening rings, should be made, this cryomodule is largely appropriate for the required cw application and provides realistic values for cryogenic loads, cavity Qs, and costing. The 12 GeV upgrade cryomodule system has well controlled microphonics and the ability for overall optimization of individual cavity performance with the low level rf control using the topology of a single klystron per accelerating structure. The 12 GeV upgrade cryomodule system supports acceleration of a total beam current of up to ~1 mA, but if necessary a coupler redesign and increased klystron power would provide a substantial increase in current capability. For a microbunch frequency of 2.5 MHz, the parameters of Table 2 would result in a total beam current per cryomodule of 1 mA given the proposed four-pass recirculation.

The recirculation topology (See Figure 1.) is similar to that of JLab with a linac, spreader, recirculation, recombination sequence and a microbunch-by-microbunch based extraction system utilizing rf separators. Different from JLab is the utilization of a single recirculated linac with separate return legs between the two arcs reducing by half the amount of beam spreading and recombining per pass and thereby reducing the potential for loss of beam quality. It is envisioned that multiple FELs will follow the microbunch-by-microbunch extraction segment with a photocathode electron gun providing microbunches optimized for individual users. The recirculated linac configuration shown in Figure 1 has an approximate footprint of ~300 x ~770 m (~57 acres).

From JLab 12 GeV upgrade data, the total SRF linac energy of 2.6 GeV of Figure 1 is compatible with two cryogenic plants each with a capacity of the largest transportable (~18 kW at 4.5K) unit for a total of ~35 kW at 4.5K. The proposed execution sequence is for Phase I to implement the full conventional facility necessary for Figure 1 but only installing 0.9 GeV of the 2.1 GeV recirculated linac. This will require only one cryogenic plant and provide output energies of approximately 1.3, 2.2, 3.1, and 4.0 GeV. (Not including the undulators, this implementation including the cost of civil construction is estimated to be comparable to the cost of an unrecirculated 4 GeV SRF linac, the appropriate cryogenic plant, and no civil construction excepting that necessary to house the cryogenic plant.) At a later date, an additional 1.3 GeV of recirculated linac and its cryogenic plant could for only an additional one third of the initial investment be straightforwardly implemented in the existing tunnel and klystron gallery providing the Phase II layout of Figure 1.

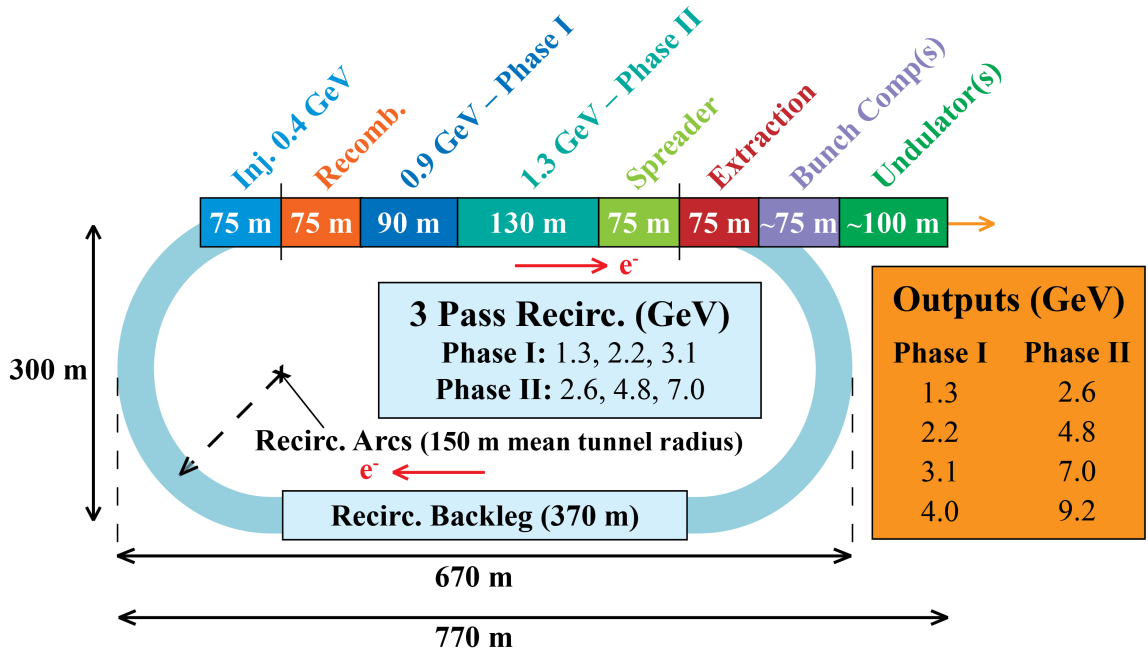

**Figure 1.** Conceptual layout of a recirculated SRF linac for a 5 keV upgradable to 25 keV FEL facility. Phase I implementation includes full civil construction and 1.3 GeV of SRF linac with cryogenic facility providing output energies of 1.3, 2.2, 3.1, and 4 GeV. Phase II implementation includes another 1.3 GeV (2.2 GeV total) of recirculated linac and an additional cryogenic plant facility providing output energies of 2.6, 4.8, 7.0, and 9.2 GeV.

The costs given assume all new construction for the recirculation topology. The footprint of Figure 1 is shown in Figure 2 along the middle third of the SLAC linac where some savings might be realized by reutilization of the existing linac tunnel and klystron gallery.

The LCLS SLAC gun is not capable of cw performance, but it does, however, provide a point of comparison for possible enhanced performance cw guns of the future. The value of $\varepsilon_N$ varies as a function of the charge per micro-bunch. In the low charge regime, the thermal dominates with the emittance scaling as the charge per bunch to the one-third power. From reference [3], the normalized emittance is given as:

$$\varepsilon_N \approx Constant \cdot \sqrt{0.111 \cdot Q^{\frac{2}{3}} + 0.18 \cdot Q^{\frac{4}{3}} + 0.18 \cdot Q^{\frac{8}{3}}} \quad \textbf{Eqn. 5}$$

Where the *Constant* is empirical (a value of 1.4 is consistent with the LCLS results), giving the normalized emittance $\varepsilon_N$ (mm-mrad) as a function of charge per bunch $Q$ (C). For a charge per bunch of 100, 200, and 300 pC, the SLAC gun provides $\varepsilon_N$ of 0.25, 0.35, and 0.43 mm-mrad respectively.

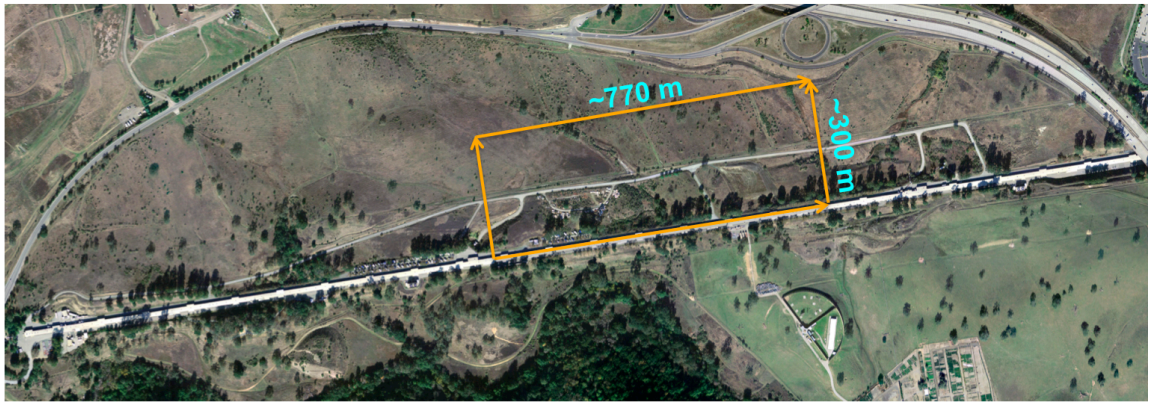

**Figure 2.** Approximate footprint of **Figure 1** along the middle one third of the SLAC linac.

For the analysis presented, the values of Table 1 were utilized for the electron beam parameters of the recirculated linac and the values of Table 2 were employed for electron beam parameters at the FEL undulator. From Eqn. 1, a planar undulator $\lambda u$ of 2 cm, and $K$ of 1 with an electron beam energy of 4 GeV or 9.2 GeV, provides $\lambda ph$ of 0.245 nm (Photon energy of 5 keV) or 0.046 nm (Photon energy of 26.8 keV). For these parameters, Eqn. 2 would require that $\varepsilon_N \sim 0.15$ and $\sim 0.07$ mm-mrad for 5 and 26.8 keV respectively. A value of 0.35 mm-mrad was the assumed for the input normalized emittance. The Ming Xie formalism [4] was used to determine the effect of finite electron beam emittance and energy spread including increases from coherent and incoherent synchrotron radiation on the length of undulator necessary to achieve saturation as given in Table 3.

**Table 1.** Electron beam parameters assumed for recirculated linac.

| Recirculated Linac Beam Parametes | Value |
|---|---|
| Input Normalized Emittance (mm-mrad) | 0.35 |
| Charge per bunch (pC) | 100 |
| Bunch Length (ps) | 2 |
| Peak Current (A) | 50 |
| Bend Radius (m) | 100 |

The implicit design assumptions made that are addressed in the Next Steps section are:

- The electron beam can be recirculated while maintaining the beam quality sufficient to support an efficient FEL process by utilizing a large (150 m) arc tunnel radius and having a modest (few degrees rf ~ 2 ps) bunch length during recirculation. A bending radius of 100 m is compatible with a gross arc radius of 150 m.

  The deleterious effects of coherent synchrotron radiation can be suppressed using techniques described by DiMitri et al. [5]. The Coherent Synchrotron Radiation (CSR) causes negligible (~0.1%) contributions to the beam emittance given a beam a bunch length of 2 ps with up to two times the charge per microbunch (200 pC).

Using a Theoretical Minimum Emittance (TME)-based recirculator arc [6], the Incoherent Synchrotron Radiation (ISR) will for the case of 4 GeV Phase I generate an increase in the normalized emittance of <0.002 mm-mrad and in rms momentum spread of $2.2 \times 10^{-6}$. Then ISR will cause no performance issues for Phase I. The ISR will for the case of 9.2 GeV Phase II, generate an increase in the normalized emittance of 0.17 mm-mrad and in rms momentum spread of $1.6 \times 10^{-5}$. [7,8,9,10]

- The bunch quality during recirculation is maintained by retaining a bunch length of order ps. As a consequence to obtain the peak current appropriate for an FEL, the electron bunch length must be reduced from ps to tens of fs by compression after acceleration. A recent publication [11], proposes a bunch compression scheme that for the case of 10 GeV electrons, rms energy spread of 500 keV, and a bunch charge of 0.2 nC provides a compression factor of 30 resulting in a peak current of 1.2 kA with a transverse emittance growth of <30%. A similar result for our case would reduce a 2 ps bunch length to ~65 fs providing a peak current of 1.5 kA as given in Table 2.

**Table 2.** Parameters assumed at the FEL undulator.

| Parameter at FEL undulator | Value |
|---|---|
| Charge per bunch (pC) | 100 |
| Bunch Length (fs) | 65 |
| Peak Current (kA) | 1.5 |
| Undulator Period (cm) | 2 |
| Undulator Parameter K (planar) | 1 |
| Electron energy spread (keV) | 500 |

Table 3 provides for the electron energies of Figure 1 and parameters of Table 1 and Table 2, the normalized emittance including the effects of ISR and bunch compression, and using the Ming Xie formalism [4], the undulator length necessary to achieve saturation.

## Next Steps

A refined parameter list including those of Table 1, Table 2, Table 3, injection energy, recirculated linac energy, recirculation bend radius, and bunch compression and seeding schemes among others can largely be quantitatively evaluated through simulations to provide a consistent and globally optimized set.

Two key simulation/experiment benchmarks are proposed. First, the effectiveness of the recirculation strategy can be judged by comparing beam measurements at a similar linac configuration (such as JLab) to simulation results to ensure efficacy of the predicted recirculation performance from simulations. Second, the appropriateness of the at-energy bunch compression strategy can be evaluated by again comparing beam measurements with simulations at an extant facility such as Jlab or SLAC.

**Table 3.** Photon energy for electron energies of **Figure 1**, and using Ming Xie formalism, the undulator length necessary for saturation given parameters of **Table 1** and **Table 2**. The normalized emittance at undulator is from (input normalized emittance plus increase from ISR) x 1.3 from bunch compression. An average machine beta function $\beta_f$ of 10 m was used for electron energies up to 4.8 GeV and 40 / 50 m for electron energy of 7.0 / 9.2 GeV respectively.

| Electron Energy (GeV) | Photon Energy (keV) | Normalized Emittance (mm-mrad) | $\rho$ Pierce Parameter | Undulator Length (m) |
|---|---|---|---|---|
| 1.3 | 0.54 | 0.46 | $11.7 \times 10^{-4}$ | 20 |
| 2.2 | 1.5 | 0.46 | $8.2 \times 10^{-4}$ | 27 |
| 2.6 | 2.1 | 0.46 | $7.4 \times 10^{-4}$ | 31 |
| 3.1 | 3.0 | 0.46 | $6.5 \times 10^{-4}$ | 36 |
| 4.0 | 5.0 | 0.46 | $5.5 \times 10^{-4}$ | 46 |
| 4.8 | 7.3 | 0.47 | $4.9 \times 10^{-4}$ | 60 |
| 7.0 | 15.5 | 0.62 | $2.2 \times 10^{-4}$ | 127 |
| 9.2 | 26.8 | 0.67 | $1.6 \times 10^{-4}$ | 206 |

## Conclusion

Given an appropriate site and a reasonable provision for conventional construction, an FEL facility based on a recirculated SRF linac meeting the recent BESAC criteria [1] of a "high repetition rate, ultra-bright, transform limited, femtosecond x-ray pulses over a broad photon energy range" can be achieved for a modest initial investment. Perhaps more importantly, the utilization of a recirculated SRF linac will provide a cost-effective opportunity for a scientifically significant, world-class FEL facility providing 25 keV photons possibly exceeding with its cw performance the scientific reach of the European XFEL.

## References

[ 1 ] http://science.energy.gov/~/media/bes/besac/pdf/Reports/Future_Light_Sources_report_BES
[2] J. Bisognano, et al., "Design Alternatives for a Free Electron Laser Facility", Proceedings of the International Particle Accelerator Conference, (2012 New Orleans LA), pgs 1777-1779. http://accelconf.web.cern.ch/AccelConf/IPAC2012/papers/tuppp079.pdf
[3] C. Pelligrini, "The Next Generation of X-ray Sources", Reviews of Accelerator Science and Technology, Vol. 3 (2010) 185-202.
[4] Xie M, *Proceedings of the 1995 Particle Accelerator Conference* (IEEE, Piscataway, NJ, 1996), p. 183. and *Nucl. Instr. and Meth. A* 455, 59
[5] S. Di Mitri, M. Cornacchia, and S. Spampinati, "Cancellation of Coherent Synchrotron Radiation Kicks with Optics Balance", Phys. Rev. Lett. 110, 014801, 2 January 2013.
[6] D. Einfeld *et al*, "A Lattice Design to Reach the Theoretical Minimum Emittance for a Storage Ring", Proc. EPAC'96.
[7] D. Douglas and C. Tennant, "Management of Synchrotron-Radiation-Driven Beam Quality Degradation During Recirculation Transport of High Brightness Electron Beams", JLAB-TN-13-036, July 19, 2013.

[8] D. Douglas et al, "ISR-Driven Energy Scaling of FEL Driver Recirculator Radius", JLAB-TN-10-048, Dec. 12, 2010.
[9] D. Douglas, "CSR Scaling: JLAMP to CEBAF-X", JLAB-TN-13-033, June 27, 2013.
[10] C. Tennant and D. Douglas, "JLAMP: Low Energy Recirculator v1.0", JLAB-TN-10-023, July 12, 2010.
[11] Yichao Jing, et al, "Compensating effect of the coherent synchrotron radiation in bunch compressors", PRSTAB, **16**, 060704 (2013).
http://prst-ab.aps.org/abstract/PRSTAB/v16/i6/e060704